\documentclass[twoside]{ilcws07}
\usepackage[latin1]{inputenc}
\usepackage[dvips]{graphicx,epsfig,color}
\usepackage{wrapfig,rotating}
\usepackage{amssymb,amsmath,array}
\def\W{\rm W}

\def\qqbar{\rm{q}\bar{\rm{q}}}
\def\Z{\rm Z}
\def\h{\rm h}
\def\ttbar{$t\bar{t}$}

\def\tbtb{$t\bar{b} \bar{t}b$}
\def\tbWm{$t\bar{b}\W^-$}
\def\tbWp{$\bar{t}b\W^+$}
\def\WWZ{$\W^+\W^-\Z$}
\def\ZZZ{\Z\Z\Z}
\def\ZZ{\Z\Z}
\def\hZ{\h\Z}
\def\hZZ{\h\Z\Z}

\def\hhZ{\h\h\Z}
\def\eeZZ{$e^+e^-\Z\Z$}
\def\enZW{$e^\pm\nu\Z\W^\mp$}
\def\llpm{\ell^-\ell^+}
\def\btag{$b$-tag}
\def\btagging{$b$-tagging }
\def\lhhh{$\lambda_{\rm hhh}$}
\def\dpflow{$\frac{\Delta E}{\sqrt{E}}$}

\begin{document}
\title{
Higgs self coupling measurement} 
\author{Djamel-Eddine BOUMEDIENE and Pascal GAY
\vspace{.3cm}\\
Laboratoire de Physique Corpusculaire - Univ. B. Pascal/IN$^2$P$^3$-CNRS \\
24 Av. des Landais, F-63177 Aubi\`ere Cedex - France
}

\maketitle

\begin{abstract}
    A measurement of the Higgs self coupling from e$^+$e$^-$
    collisions in the International Linear Collider is presented. The
    impact of the detector performance in terms of $b$-tagging and
    particle flow is investigated.

\end{abstract}

\section{ Introduction }
The trilinear Higgs self coupling, \lhhh, is extracted from
the measurement of the cross-section, $\sigma_{\rm hhZ}$, of the Higgs-strahlung process
$e^+e^- \rightarrow \Z\h\h$\cite{Theo}.
This study is performed in the standard model
framework assuming $m_{h}=120~\rm{GeV}/\rm{c}^2$ at $\sqrt{s}=500~\rm{GeV}$. At this
centre-of-mass energy the
{\W} fusion process ($e^+e^- \rightarrow \nu\bar{\nu} \h\h$ in t-channel) is
negligible. It has been therefore established that
$\frac{\Delta\lambda_{\rm hhh}}{\lambda_{\rm hhh}} \simeq 1.75
\frac{\Delta\sigma_{\rm \hhZ}}{\sigma_{\rm \hhZ}}$. 
All the results are given for a luminosity of $2~\rm{ab}^{-1}$.

\section{ Monte Carlo simulation }
\label{sec:montecarlo}

 The signal and background event samples have been generated with {\tt
 Whizard}\cite{whizard}. {\tt PYTHIA}\cite{pythia} has been used to perform the
 hadronisation of the primary partons. 
  Table~\ref{tab:mc} summarizes the cross-section of the simulated processes.
 At $\sqrt{s}=500$~GeV the dominant background processes involve top
 quarks. They are simulated, as well as final states with two and
 three bosons. 305 hhZ events are expected for an estimated background
 three orders of magnitude above.

 The detector is simulated through a parametric Monte
 Carlo,~\cite{simdet} in which 
 the subdetectors are characterized by their acceptance
 angles, resolutions and energy thresholds. The intrinsic energy resolutions,
 $\frac{\Delta E}{\sqrt{E}}$, of the electromagnetic
 (ECAL) and hadronic (HCAL) calorimeters are respectively of 10.2\% and 40.5\%.

 The \btagging efficiency and $c$-jet contamination are
 parametrized according to the full reconstruction~\cite{hawkings}. In
 this study, a \btagging efficiency, $\epsilon_b$, of $90\%$ has been
 chosen, value which is not necessarily the best working point (cf. section~\ref{sec:btag})
\begin{table}[htbp]
  \centering
  \begin{tabular}[]{|c||c|c|c|c|c|c|c|c|c|c|}
    \hline 
   \small Final state & \small\hhZ & \small\hZ & \small\hZZ & \small\ZZ &  \small\ZZZ &\small\WWZ & \eeZZ
   &\small\enZW
   \\ \hline
   \small $\sigma$ (fb) & \small 0.1528  &\small 14.1 &\small 0.5 &\small 45.12 &\small 1.05 &\small 35.3 &\small 0.287
   &\small 10.09\\ \hline
   \small Nb. events &\small 20k &\small 110k &\small 10k &\small 110k &\small 20k &\small 130k &\small 10k &\small 60k \\
   \hline \hline
   \small Final state & \small\ttbar &\small\tbWp,\tbWm
   &\small\tbtb&\small \ttbar\Z &\small \ttbar h &\small \ttbar$\nu\bar{\nu}$ & $\nu\bar{\nu}\ZZ$
   &\small$\nu\bar{\nu}\W^+\W^-$ \\ \hline
   \small $\sigma$ (fb) &\small 526.4 &\small 16.8 &\small 0.70 &\small 0.6975 &\small 0.175 &\small 0.141 &\small 1.083
   &\small 3.627 \\ \hline
   \small Nb. events &\small 1M &\small 240k &\small 20k &\small 20k &\small 20k &\small 20k &\small 20k &\small 30k \\
   \hline
  \end{tabular}
  \caption{\small Cross sections of the simulated processes 
  and number of generated events.}
 \label{tab:mc}
\end{table}

For each event the boson masses are reconstructed according to a
final state hypothesis. The $b$-content of the event is obtained from an 
estimation of the number of the $b$-like jets in the event.

\section{ Event Selection and cross section measurement }
\label{sec:mesure}

The hhZ final state is sorted into three channels that correspond to the
three Z decay modes $\Z \rightarrow \qqbar$, $\Z \rightarrow \nu\bar{\nu}$ and $\Z
\rightarrow \llpm$. In order to define the three samples
representing these three channels, a preselection is applied on the signal and
the background events, based on the following criteria: 

\begin{wrapfigure}{r}{0.5\columnwidth}
\vspace{-0.6cm}
\centerline{\includegraphics[width=0.45\columnwidth]{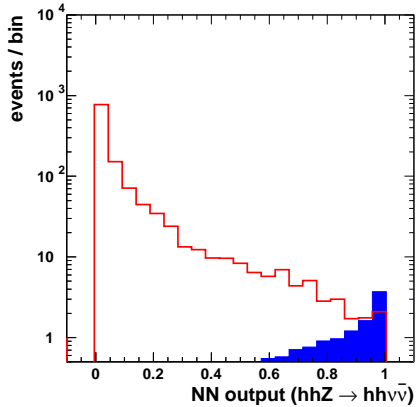}}
\caption{\small Distribution of the Neural Network defined for the hh$\nu\bar{\nu}$ channel
  after preselection. The plain histogram represents the signal
  contribution while the line represents the background contribution.}\label{fig:nnplot}
\vspace{-1.2cm}
\end{wrapfigure}

\begin{itemize}
\item Global $b$ content : only events with a minimal $b$-content are
  selected.   The criteria value used to select hh$\qqbar$ events (six jets
  topology) is different from the one used to select 
  hh$\nu\bar{\nu}$ or hh$\llpm$ events (four jets topology),

\item  the visible energy is used to define two exclusive samples
  which correspond to hh$\nu\bar{\nu}$ events (visible energy below
  $0.75 \sqrt{s}$) and hh$\qqbar$ + $\h\h\llpm$
  events (visible energy higher then
  $0.75 \sqrt{s}$),

\item the identification of two isolated leptons compatible with a Z boson
  mass allows to separate hh$\llpm$ from hh$\qqbar$ events.
\end{itemize}

 \noindent Two variables are used to select the hhZ final state : a
 Neural Network (NN)~\cite{nnpaw} and the global event \btag.
 Three kinds of inputs are used to feed the NN:
\begin{itemize}
\item event shape variables as, for instance, charged multiplicity, sphericity or
thrust values,
\item combinations of the different di-jet masses assuming a given final state,
\item global $b$-flavor content of the event.
\end{itemize}

A Neural Network is designed for each of the hh$\qqbar$,
hh$\nu\bar{\nu}$ and hh$\llpm$ final states.
The neural networks are trained on large preselected samples of simulated events
including all expected processes listed in Section~\ref{sec:montecarlo}.
The output of the neural network for the hh$\nu\bar{\nu}$ selection is displayed on
Figure~\ref{fig:nnplot}.

 For each channel, the cuts on NN and global \btag\space are defined in order to maximize the
 figure of merit $\delta=s/\sqrt{s+b}$. 
 The combination of the three selections leads to 128 events expected
 from the background processes considered and
72 events from hhZ process corresponding to a $\delta$ value of 5.2.

 In order to extract the cross section of the hhZ creation process, 
a Likelihood maximization method is used. It is based on the two dimensional 
NN $\times$ \btag\space distribution.
The expected precision on the cross-section measurement is 16\%.
Therefore, the expected precision on \lhhh\space is 28\%. 
This result is
obtained for a particle flow resolution of $\frac{30\%}{\sqrt{E}}$ and a
\btag\space of 90\%. A better working point for the \btag\space efficiency may
be found, as it will be shown in next section.

\section{Scan of the detector performance parameters}

 Two parameters have been investigated : the particle flow resolution and the \btagging.
 The full analysis described in section~\ref{sec:mesure} has been
 performed and optimized for each hypothesis on the detector performance.
 The selection has been performed with different Neural Networks which
 combine the same input variables with adapted
weights and re-optimized cuts.

\begin{wrapfigure}{r}{0.5\columnwidth}
\vspace{-1.cm}
\centerline{\includegraphics[width=0.45\columnwidth]{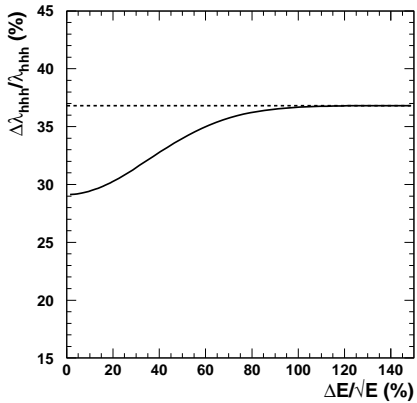}}
\caption{\small Expected resolution on \lhhh\space as a function of
  the particle flow resolution. $\epsilon_b$ is fixed to 90\%.}
\label{fig:efscan}
\vspace{-0.5cm}
\end{wrapfigure}

\subsection{Particle flow impact on the measurement}

\label{sec:pflow}
The particle flow uncertainty influences the jet pairing (based
on di-jet masses) and propagates to
the energies and momenta of the
reconstructed bosons and then to the selection inputs. Then different
efficiencies of the event selection are observed.

A fast simulation is used in order to determine the impact of the
particle flow resolution on $\Delta\lambda_{\rm hhh}$.

For each event, the stable and visible particles (i.e. all stable
except neutrinos) are considered with their generated energies and
momenta with no detector simulation. They are clusterized in jets. A
jet by jet smearing of the calorimeter cluster energies is then applied in
order to simulate the combined effect of the detector
resolution and the particle flow algorithms.

This study investigates the direct impact of the calorimetric
resolution and the particle flow algorithms on the precision independently of its impact on the
jet clusterisation. A $\frac{\Delta E}{\sqrt{E}}$ resolution range
from $0\%$ to $130\%$ has been covered.

\subsection { Event simulation with various \btag\space efficiencies }
\label{sec:btag}
 The measurement performance depends also on the \btag. For each jet as defined in section~\ref{sec:pflow} consisting on $b$-fragmentation products,
a \btag\space is statistically defined assuming a given
efficiency $\epsilon_{b}$. For a given vertex detector (VDET) and a given
jet energy, the values of $\epsilon_b$ is associated to a $c$ flavored contamination
(quantified by $\epsilon_c$), namely the rate of $c$-jets identified as
$b$-jets. Similary a rate of $uds$-jets identified as $b$-jets is
associated to $\epsilon_b$ and it has also been taken into account.
$\epsilon_b$ was varied in the range $40\%$ to $95\%$.

\subsection { Results }

 The dependence of the precision on the measurement of \lhhh, with respect to
the particle flow uncertainty is displayed on Figure~\ref{fig:efscan}.

For a given \btag\space efficiency, the uncertainty on the \lhhh\space
measurement increases when \dpflow\space increases.
For $\epsilon_{b}=90\%$, the best measurement is $29\%$ when a perfect particle flow is
assumed while for higher resolution on particle flow the precision increases to
$37\%$. The improvement of the particle flow enhances the precision on the trilinear
coupling by a factor $1.3$. This gain is equivalent to a factor $1.7$ on the
required luminosity.

\begin{wrapfigure}{r}{0.5\columnwidth}
\vspace{-1.1cm}
\centerline{\includegraphics[width=0.45\columnwidth]{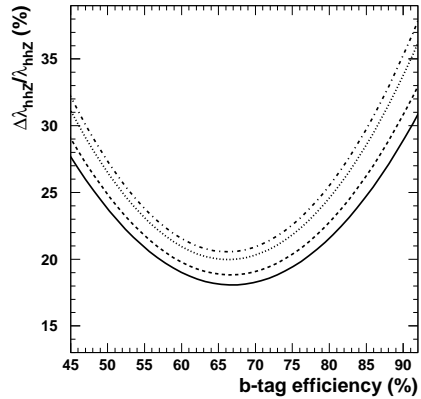}}
\caption{\small Expected resolution on the \lhhh\space as a function of
 $\epsilon_b$. From down to up $\frac{\Delta E}{\sqrt{E}}$=0\%, 30\%,
 60\% and 130\%.}\label{fig:btgscan}
\vspace{-1.cm}
\end{wrapfigure}

 The dependence of the precision on the Higgs self coupling measurement
with respect to the \btag\space efficiency is displayed on
Figure~\ref{fig:btgscan} where an optimum is observed around
$\epsilon_{b}=67\%$. This \btag\space efficiency corresponds,
for a typical jet energy of 45~GeV, to
$\epsilon_c \simeq 3\%$ which means 
that the \hhZ\space final state measurement is optimized for pure \btagging.

\section{Conclusion}
The feasibility of the \lhhh\space measurement was established. The
expected statistical precision with a typical detector is about 28\%.
It was shown that an optimization of the \btagging\space allows to reduce
this uncertainty to 19\%.

\begin{footnotesize}

\end{footnotesize}

\begin{thebibliography}{99}
\bibitem{url} Slides: \\
\verb$http://ilcagenda.linearcollider.org/contributionDisplay.py?contribId=155&sessionId=71&confId=1296$
\bibitem{Theo} A.~Djouadi, W.~Killian, M.~Muhlleitner and P.~Zerwas,
  Eur. Phys. J. {\bf C10}
\bibitem{whizard} W.~Kilian, T.~Ohl, J.~Reuter, WHIZARD: Simulating Multi-Particle Processes at LHC and ILC, arXiv: 0708.4233 [hep-ph]
\bibitem{pythia} T.~Sjostrand, S.~Mrenna and P.~Skands JHEP05 (2006) 026.
\bibitem{simdet} M.~Pohl, H.~J.~Schreiber, SIMDET.3 A Parametric Monte
                 Carlo for a TESLA Detector, DESY 99-030 (1999)
\bibitem{hawkings} R.~Hawkings, Vertex detector and flavour tagging
  studies for the Tesla linear collider, LC-PHSM/2000-026
\bibitem{nnpaw} J.~Schwindling, MLPFit: A tool for Multi-Layer
  Perceptrons,\verb$ http://home.cern.ch/~schwind/MLPfit.html$
\end{thebibliography}
\end{document}